# The high hydrogen adsorption rate material based on graphane decorated with alkali metals

Liubov Yu. Antipina [1], Pavel V. Avramov [2], Seiji Sakai [2], Hiroshi Naramoto [2], Manabu Ohtomo [2], Shiro Entani [2], Yoshihiro Matsumoto [2], and Pavel B. Sorokin* [1,3]

[1] *Technological Institute of Superhard and Novel Carbon Materials, 7a Centralnaya Street, Troitsk, Moscow region, 142190, Russian Federation*

[2] *Advanced Science Research Center, Japan Atomic Energy Agency, 2-4 Shirakata Shirane, Tokai-mura, Naka-gun, Ibaraki-ken 319-1195, Japan*

[3] *Emanuel Institute of Biochemical Physics of RAS, 119334 Moscow, Russian Federation*

The graphane with chemically bonded alkali metals (Li, Na, K) was considered as potential material for hydrogen storage. The ab initio calculations show that such material can adsorb as many as 4 hydrogen molecules per Li, Na and K metal atoms. These values correspond to 12.20 wt%, 10.33 wt% and 8.56 wt% of hydrogen, respectively and exceed the DOE requirements. The thermodynamic analysis shows that Li-graphane complex is the most promising for hydrogen storage with ability to adsorb 3 hydrogen molecules per metal atom at 300 K and pressure in the range from 5 to 250 atm.

## I. INTRODUCTION

The biggest challenge in the application of pollution-free hydrogen engine is the realization of a storage material which can effectively adsorb and desorb the volatile and explosive hydrogen gas. Graphene-based nanostructures are considered as a promising hydrogen adsorber due to the low specific mass and large surface area. In recent decades a considerable number of papers have been devoted to the investigation of carbon nanotubes (CNT), [1,2] graphene [3,4] and fullerene [5] as a possible hydrogen storage material. Carbon nanostructures (including graphene) display too small binding energy of hydrogen molecules (~ 0.05 eV per molecule) whereas for the effective application it requires the binding energy in the range of ~ 0.20 … 0.40 eV per molecule. On the other side, carbon nanostructures with bonded metal atoms (organometallic complexes) show high adsorption energy (~ 0.20 … 0.60 eV), and carbon nanotubes, [6,7] graphene, [8,9] carbon clusters, [10] and fullerene $C_{60}$ [11,12] decorated with alkali (Li, Na, K) [6,9,10,13-16] and alkali-earth (Ca) [7,8,12,17,18] metal atoms are currently proposed as promising hydrogen storage materials. Particularly, it has been shown [6] that single-walled carbon nanotubes doped with alkali metals (Li and K) can adsorb 14 wt% (Li) and 20 wt% (K) of hydrogen at moderate conditions, in contradiction with lower values reported later. [13] The activated carbon doped with lithium [19] can store from 2.10 to 2.60 wt% of hydrogen at 77 K and at 2 MPa $H_2$. Fullerenes [16,20] doped with Li has been shown to adsorb 0.47 wt% at 77 K and 2 atm and 2.59 wt% at 523 K and 3 MPa, respectively. Also it has been reported that the lithium-doped metal-organic framework impregnated with lithium-coated fullerenes can adsorb 5.1 wt% (298 K, 100 atm) and 6.3 wt% (243 K, 100 atm), [21] respectively.

In such organometallic complexes the hydrogen molecules are adsorbed mainly by metal atoms whereas carbon nanostructures play the role of the base. So, the reduction of relative amount of carbon atoms increases the total hydrogen adsorption rate. The carbon base structure should suppress the metal aggregation. It was shown [22,23] that a low barrier of metal atoms diffusion on a carbon nanotube leads to the metallic aggregation along with dramatic decrease of the hydrogen adsorption rate. Such problem can be solved by introducing of defects in the structure [7,24] or using the ultrathin structures e.g. carbyne. [25]

Graphane which was predicted [26] and synthesized [27] quite recently satisfy given above requirements. Graphane's crystalline lattice consists of hexagonal carbon net in which every carbon atoms bounded with hydrogen in chair configuration. Graphane can be considered as a thinnest diamond film due to pure $sp^3$ hybridization of all carbon-carbon bonds and absence of conductive π-bands.

In this paper, the results of investigation of organometallic complexes of graphane with alkali metals (lithium, Li-Gr, sodium, Na-Gr and potassium, K-Gr) as a possible hydrogen storage media are presented. In the considered materials a part of the graphane's hydrogen atoms are substituted by metal atoms. The problem of undesirable metal aggregation is eliminated in such materials because every carbon atom is bounded either with metal atom or hydrogen atom which prevents the metal diffusion. It should be noted that in the experimental studies partially hydrogenated graphene (graphane) was obtained which was justified by semiconducting behavior of conductivity of the fabricated material [27] (instead of predicted insulating behavior) [26] and a theoretical simulation of graphene hydrogenation process. [28] The non hydrogenated carbon atoms allow to adsorb the metal atoms and form the considered organometallic complex.

The analysis of various metal arrangements and concentrations on graphane was carried out and energetically favorable metal-graphane (Me-Gr) complexes with enough binding energy for the hydrogen storage were found. It was obtained that at zero temperature every metal atom in the Me-Gr complex can adsorb up to 4 hydrogen molecules with binding energy ~ 0.20 eV. The theoretical limits of hydrogen storage amount for Li-Gr, Na-Gr and K-Gr are estimated as much as 12.20 wt%, 10.33 wt% and 8.56 wt%, respectively. These values satisfy DOE requirement for commercial use of hydrogen in transport. The investigation of adsorption thermodynamics in Me-Gr complexes shows that Na-Gr with adsorbed four hydrogen molecules should be stable at low temperature (T ≤ 250 K). K-Gr can adsorb up to 2 $H_2$ molecules at the pressure 100 atm and temperature 300 K, the adsorption of more hydrogen molecules requires more tough conditions. Li-Gr looks the most promising material for the hydrogen storage because one lithium atom can adsorb 3 hydrogen molecules at T = 300 K in the pressure ranges 5 - 250 atm which corresponds to 9.44 wt% hydrogen adsorption capacity.

## II. COMPUTATIONAL DETAILS

All calculations of atomic and electronic structure of Me-Gr 2D nanostructures were performed by Quantum Espresso package [29] at DFT level of theory in local density approximation using Perdew-Zunger parameterization [30] and plane wave basis set. The plane-wave energy cutoff was equaled to 30 Ry. To calculate equilibrium atomic structures, the Brillouin zone was sampled according to the Monkhorst–Pack [31] scheme with a 8×8×1 $k$-point convergence grid. To avoid interaction between neighboring Me-Gr layers, the translation vector along *c* axis of hexagonal supercells was greater 15 Å. For validation of the chosen approach the Li-graphene, Na-graphene and K-graphene binding energies were calculated and -1.1 eV, -0.9 eV and -0.6 eV were obtained for Li, Na and K, respectively. These data are in good agreement with the reference data of -1.6 eV,[14] -0.7 eV[32] and -0.9 eV,[33] respectively.

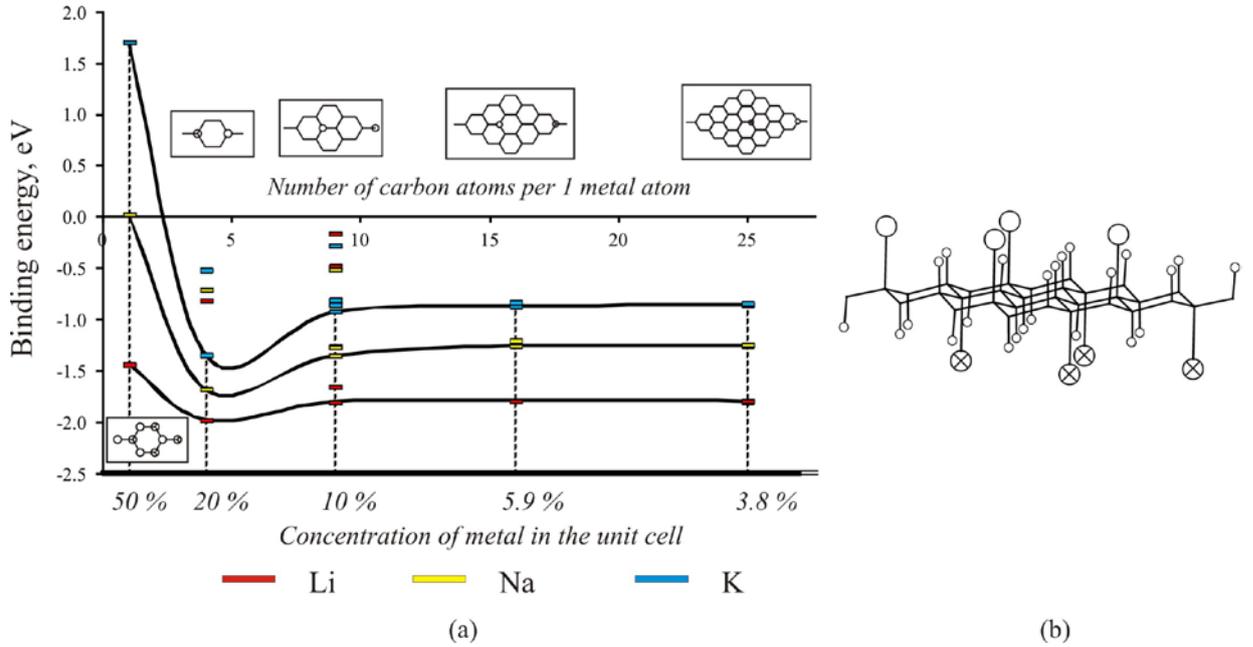

Fig. 1. (a) The dependence of binding energy of alkali metal atoms with graphane upon concentration. For each metal concentration the largest values of binding energy are connected by the solid curve, the correspondent structures are shown in the insets; (b) The perspective view of energy favorable configuration of Me-Gr complex is shown for metal concentration 20%. Carbon atoms are drawn by lines, metal atoms are depicted by circles (empty and filled circles are corresponded to outer and inner bounded metal atoms, respectively) and hydrogen atoms are shown by circles with smaller size.

### III. RESULTS AND DISCUSSION

The binding energy of metal atom on graphane $E_{bind}(Me)$ was calculated using the following expression:

$$E_{bind}(Me) = (E_{C-Me} - nE_{Me} - E_C)/n, \qquad (1)$$

where $E_{C-Me}$, $E_{Me}$, and $E_C$ are total energies of Me-graphane complex, single metal atom and graphane without *n* hydrogen atoms, respectively; and *n* is a number of metal atoms in the unit cell. The dependence of $E_{bind}$ upon the concentration of metal atoms in the various arrangements is shown in Fig. 1(a).

Me-Gr complexes with 20% metal concentration (Fig. 1(b)) are energetically favorable for all species. The pronounced binding energy minimum in all cases can be explained from the geometric point of view. The Me-Gr complex with 20 % metal concentration is energetically favorable because the distances between neighboring metal atoms $d_{Me-Me}$ (~4.5 Å) are close to the bulk values which are equaled to 3.03 Å, 3.72 Å and 4.54 Å[34] for the Li, Na and K bulk crystals, respectively. The increase of the metal concentration to 50% decreases the metal-metal distance (up to 2.5 Å), increases the repulsive forces and lowers the energetic stability of Me-Gr complex. The energy of the lithium atom on graphane for 50% concentration (-1.44 eV, each carbon atom is bonded with Li) is in accord with the work of Yang [35] (-1.40 eV). In the case of Na-Gr and K-Gr structures the strain of the metal-metal bond is too high (bonds are in 1.5 and 1.8 times smaller than corresponding bond values for Na and K bulk crystals, respectively) and the complex becomes energetically unstable which also corresponds to reference data. [15] The lowering of the metal concentrations from the 20% leads to the rapid decreasing of binding energy to -1.80 eV, -1.25 eV and -0.86 eV for lithium, sodium and potassium, respectively due to the breaking the metal-metal bonds which do not contribute in the binding energy anymore: at 10 % metal concentration the $d_{Me-Me}$ equals to ~7.5 Å which in 2.5, 2.0 and 1.7 times larger than corresponding bond values for Li, Na and K bulk crystals. It should be noted that the metal-metal distance at the 20 % concentration is closest to the potassium bulk value which is reflected on the largest energy minimum in K-Gr potential curve in comparison with other considered metal-graphane complexes (see Fig. 1(a)). The values of binding energy for such concentration are in the range of -1.30 eV ... -2.00 eV, which indicate a high energetic stability comparable with the binding energy of lithium, [14] sodium [32] and potassium [33] on graphene.

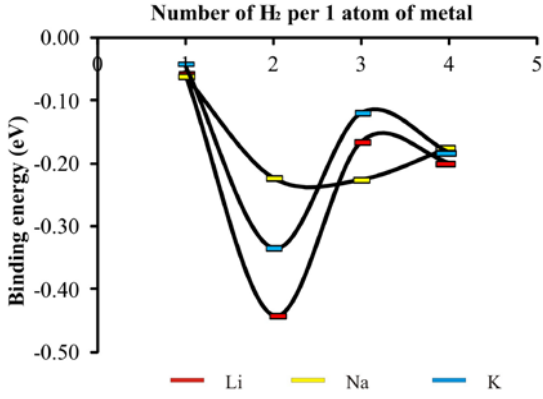

Fig. 2. The dependence of binding energy of $n^{\text{th}}$ hydrogen molecule on the Me-Gr complex upon the number of adsorbed $H_2$ molecules per the metal atom.

For the Me-Gr with 20% metal concentration, the binding energy $E_{nH_2}$ of an adding of $n^{\text{th}}$ hydrogen molecule to Me-Gr complex with already adsorbed $n-1$ $H_2$ was calculated according to following expression:

$$E_{nH_2} = \tfrac{1}{2}\left(E_{nH_2}^{Me-Gr} - E_{(n-1)H_2}^{Me-Gr} - 2E^{H_2}\right) \quad (2)$$

where $E_{nH_2}^{Me-Gr}$ and $E_{(n-1)H_2}^{Me-Gr}$ are the total energies of the complex with adsorbed $n$ and $n-1$ hydrogen molecules per metal atom, respectively; $E^{H_2}$ is the $H_2$ energy. The last term is doubled because each unit cell contains two metal atoms on both sides of graphane plane adjusted with one $H_2$ molecule each. In Fig. 2 the dependence of the binding energy upon the number of adsorbed hydrogen molecules per metal atom is displayed and in Table I the values of $E_{nH_2}$ and gravimetric capacity for each complex are presented. The data demonstrate that the capacity of Me-Gr complex exceeds the DOE target of 6.5 wt% even in the case of every metal atom adsorbs three hydrogen molecules. The increase of the number of adsorbed $H_2$ molecules up to four per metal atom leads to decreasing of binding energy up to -0.20 eV. This value is still enough for the effective storage.

Table I. Additional binding energy and hydrogen adsorption capacities of the Me-Gr complexes.

| Number of hydrogen molecules per 1 metal atom | Lithium | | Sodium | | Potassium | |
|---|---|---|---|---|---|---|
| | wt% | $E_{bind}$ (eV) | wt% | $E_{bind}$ (eV) | wt% | $E_{bind}$ (eV) |
| 1 | 3.36 | -0.06 | 2.69 | -0.06 | 2.21 | -0.04 |
| 2 | 6.50 | -0.44 | 5.30 | -0.22 | 4.37 | -0.33 |
| 3 | 9.44 | -0.17 | 7.85 | -0.23 | 6.49 | -0.12 |
| 4 | 12.2 | -0.20 | 10.33 | -0.17 | 8.56 | -0.18 |

In Fig. 3 the relaxed atomic geometries of the Me-Gr complexes with adsorbed hydrogen molecules are presented. First adsorbed H$_2$ (Fig. 3(a)) locates on the top of metal atom in all cases. The increase of the metal atom radius leads to the consequent increase of the distance between Me and H$_2$ (1.96 Å, 2.48 Å and 2.90 Å for lithium, sodium and potassium, respectively). The next hydrogen molecules arrange in the lateral directions due to hybridization effect described below.

The character of hydrogen molecule adsorption on alkali atoms depends on the metal type. Since the atomic radius of lithium is small, the first three adsorbed hydrogen molecules arrange around the ion at the distance of 1.92 - 2.26 Å (Fig. 3(a-c)) whereas in the sodium and potassium cases the H$_2$ distribute between all atoms on the distances 2.40 - 2.66 Å and 2.75 - 2.90 Å for K-Gr and Na-Gr complexes, respectively which leads to the symmetrical arrangement of the hydrogen molecules in a hexagonal order around the metal. The increase of hydrogen concentration leads to shortening Me-H$_2$ distance and consequent increase of bond lengths between metal and carbon atoms.

In the Li-Gr case the forth H$_2$ molecule settles down between two metal atoms because the atomic radius of lithium is too small to hold 4 molecules of hydrogen. The atomic radius of potassium is about 1.5 times larger the lithium one, therefore two adsorbed hydrogen molecules arrange in hexagonal order in K-Gr complex (Fig. 3(b)). The fourth H$_2$ molecule arranges on the top of K ion and displays the smaller binding energy than $E_{3H_2}$ in contrast to Li-Gr case. Distances between H$_2$ and potassium ion are in the range of 2.75 - 2.90 Å and in contrast to Li-Gr complex, the increase of hydrogen concentration leads to the shortening of K-H$_2$ distance to 2.75 Å whereas in the case of lithium this distance increases up to 2.06 Å

The sodium displays intermediate properties between lithium and potassium. The first two H2 molecules locate nearby the Na ion similar to Li-Gr case (Fig. 3(a,b)) but the next H2 molecules arrange in hexagonal order like in the case of K-Gr (Fig. 3(c,d)). The Na-H2 distances are about ~ 2.40 - 2.66 Å in all cases which is larger then Li-H2 ones but smaller than K-H2 distances. Bond lengths between metal and carbon atoms increase with increasing of hydrogen concentration: from 2.02 Å to 2.06 Å for Li, from 2.34 Å to 2.40 Å for Na, and from 2.65 Å to 2.75 Å for K.

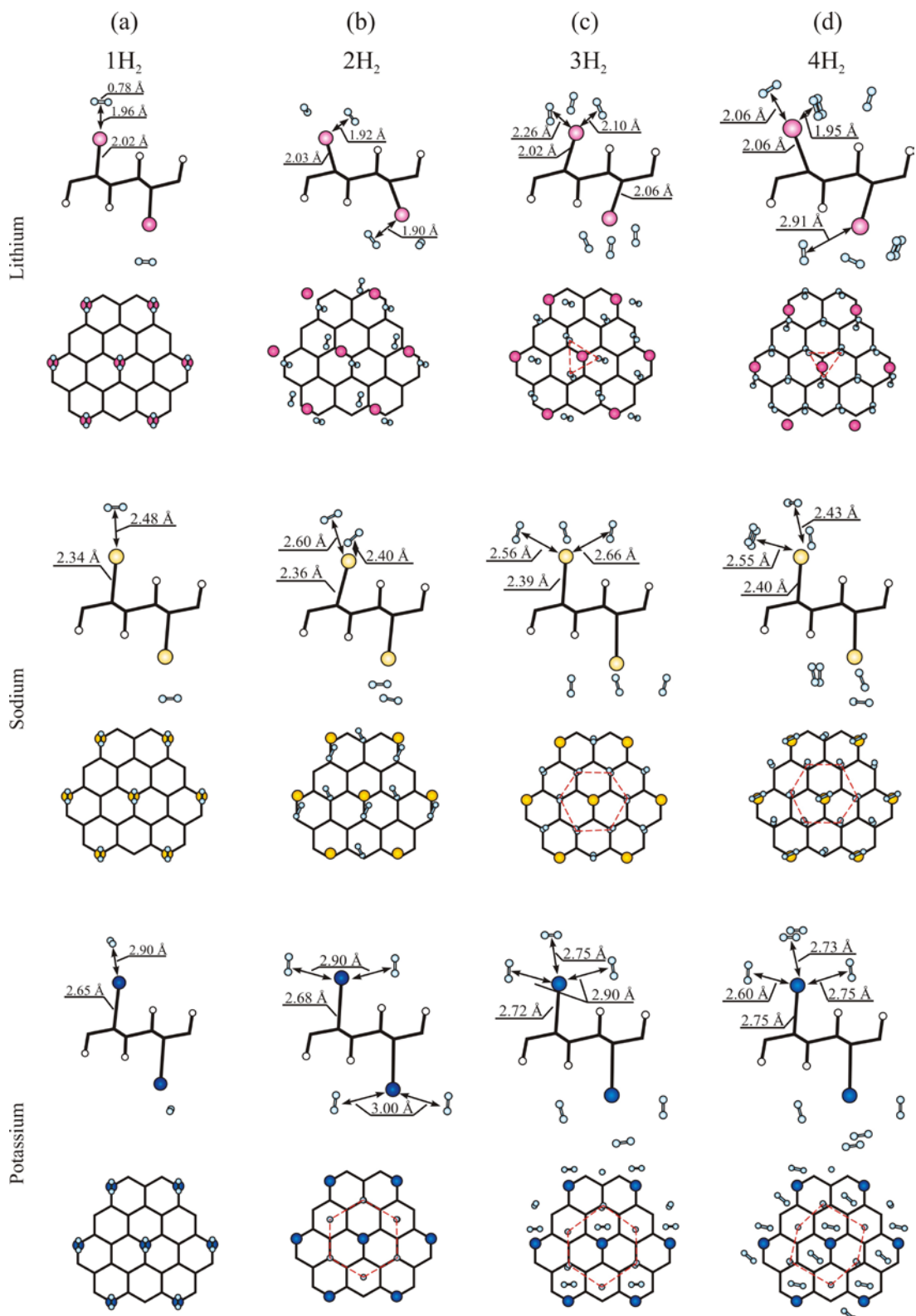

**Fig. 3.** Top and side views of atomic structures of Me-Gr complexes with one (a), two (b), three (c) and four (d) adsorbed hydrogen molecules per metal atom. Carbon atoms are drawn by lines, metal atoms are marked by circles of larger radius and filled by red, yellow and blue colors for lithium, sodium and potassium atoms, respectively. H atoms are marked by cyan and white circles with smaller radius for the molecular and atomic hydrogen forms, respectively. The triangular (in the Li case) and hexagonal (in the Na and K cases) arrangements of adsorbed hydrogen molecules are depicted by red dashed lines.

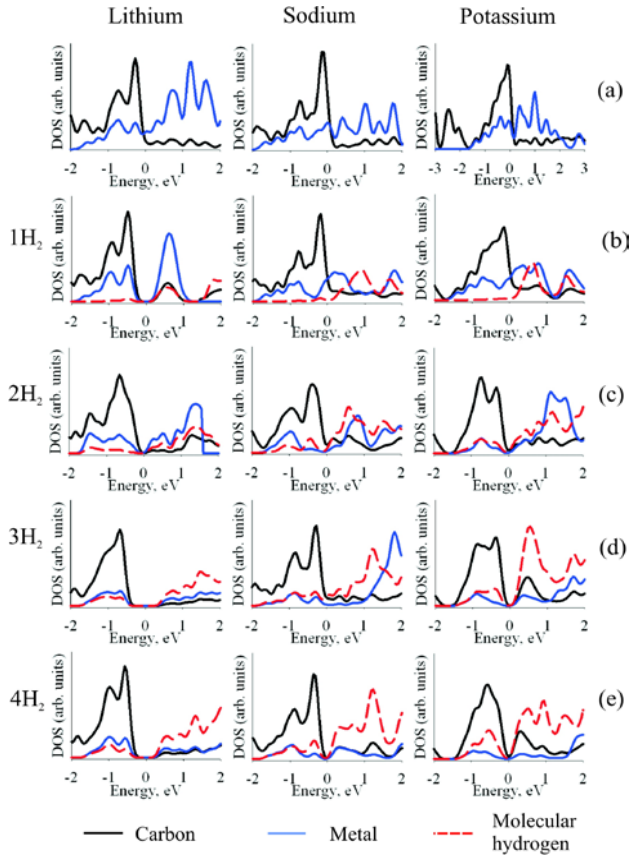

**Fig. 4.** Density of electronic states of pure (a) Me-Gr complexes and with adsorbed one (b), two (c), three (d) and four (e) hydrogen molecules per metal atom.

The densities of states (DOS) of the Me-Gr structures are presented in Fig. 4. Binding of the metal atom with graphane leads to the closing of graphane band gap due to the mixing of Me and carbon orbitals. Alkali metals transfer the charge to the carbon ($\delta$Li = +0.22, $\delta$Na = +0.46, $\delta$K = +0.46 from Löwdin population analysis) due to the smaller electronegativity ($X_{Li}$ = 0.912, $X_{Na}$ = 0.869, $X_K$ = 0.734, $X_C$ = 2.544).[36] As a result, the Me-Gr complexes display the dipole moments and polarize the neutral hydrogen molecules.[37] The first $H_2$ binds to the top of Me atom, with hybridization of $H_2$ and Li $p_z$ orbitals[37] whereas the next hydrogen molecules arrange in the lateral directions due to the hybridization of hydrogen orbitals with $p_x$ and $p_y$ orbitals of Me.

Due to molecule polarization, the first hydrogen molecule binds to all considered metal ions by electrostatic interaction.[9,38] The shape of the DOS (Fig. 4(b)) shows that the hydrogen orbitals do not participate in the formation of bonding molecular orbitals near the Fermi energy whereas the adsorption of the next hydrogen molecules leads to the overlapping of metal and $H_2$ states (Fig. 4(c-e)) which explains the observed increasing of binding energy (see Fig. 2). Such interaction was described in a number of papers [39-41] and was named in [39] as "anti-Kubas" interaction. The band gap opens after adsorption of the first, second and fourth hydrogen molecules for all Me-Gr complexes. An addition of the next hydrogen molecules causes the graduate increasing of the width of band gap except the case of lithium-graphane complex in which the adsorption of the second hydrogen molecule slightly decreases it. The relative stability of Me-Gr composites can be estimated using the following equation: [42-45]

$$E_s = E_{nH_2}^{Me-Gr} - E^{Me-Gr} - nE^{H_2} - n\mu_{H_2}(T,P) \qquad (3)$$

where $\mu_{H_2}(T,P)$ is the chemical potential of hydrogen molecular gas at given temperature and partial $H_2$ pressure which can be calculated by formula:

$$\mu_{H_2}(T,P) = \Delta H - T\Delta S + kT\ln\left(\frac{P}{P^0}\right) \qquad (4)$$

where $\Delta H$ and $\Delta S$ are the differences of enthalpy and entropy ($P^0$ = 1 atm) of given and zero temperatures which values was obtained from the reference table.[46] In Fig. 5 the dependencies of Gibbs energies $E_s$ upon the chemical potential $\mu_{H_2}(T,P)$ (and partial pressure of molecular hydrogen gas at various temperatures) are shown. Increasing of the chemical potential, which is equaled to increasing of gas pressure, leads to increasing of the stability of the system. At 0 K, all structures with adsorbed hydrogen molecules display negative binding energy but, with increasing of the temperature at low pressures, the energy becomes positive and thus at standard conditions (1 atm, 298 K), the structures are unstable. However, the adjusting of the temperature and pressure allows one to make the adsorption energy favorable.

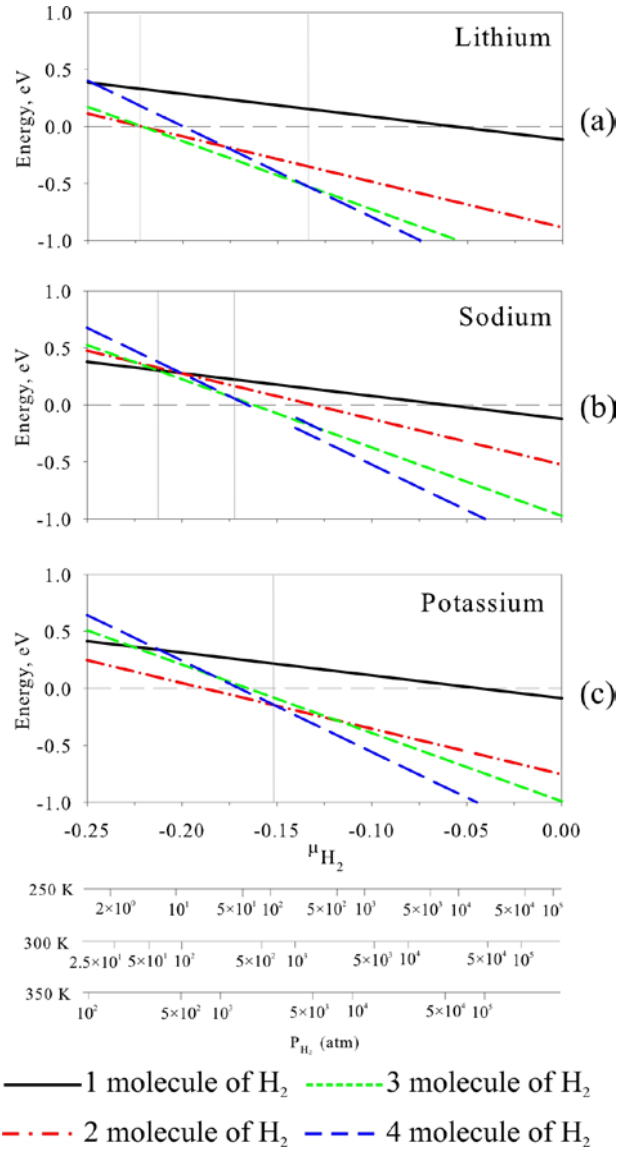

Fig. 5. The Gibbs energies of adsorption of hydrogen molecules upon the Li-Gr (a), Na-Gr (b) and K-Gr (c) complexes as the function of $H_2$ chemical potential. The positive values of the energy corresponds to hydrogen desorption. The alternative axes show the pressure of molecular hydrogen gas corresponding to chemical potential at 250, 300 and 350 K.

The Li-Gr complex adsorbs three hydrogen molecules at sufficiently mild conditions (up to 100 atm at T = 300 K) (Fig. 5(a)). At higher pressures ($\mu \approx -0,13$) the complex absorbs the fourth hydrogen molecule. The increasing of the temperature at constant pressure makes the structure unstable which means the spontaneous hydrogen desorption.

In the case of sodium at high pressure (P > 300 atm, T = 300 K), the Na-Gr complex adsorbs four $H_2$ molecules, bypassing the configuration with three or less adsorbed hydrogen molecules (Fig. 5(b)). At lower temperature (T ≤ 250 K) such complex adsorbs four hydrogen molecules (10.33 wt%) at pressure larger 55 atm. The increase of the temperature leads to structure instability and hydrogen desorption.

The K-Gr complex can adsorb either two or four $H_2$ molecules. Two $H_2$ molecules are adsorbed at pressures larger 154 atm at T = 300 K or at lower pressures (20 atm) and smaller temperature T = 250 K (Fig. 5(c)). Heating to T = 350 K leads to the decrease of the adsorption rate and adsorption process becomes favorable at very high pressures (above 750 atm), which is not always achievable.

The significant dependence of the adsorption ability of Me-Gr complex upon the temperature and pressure allows one to design a device with effectively controlled hydrogen sorption-desorption process.

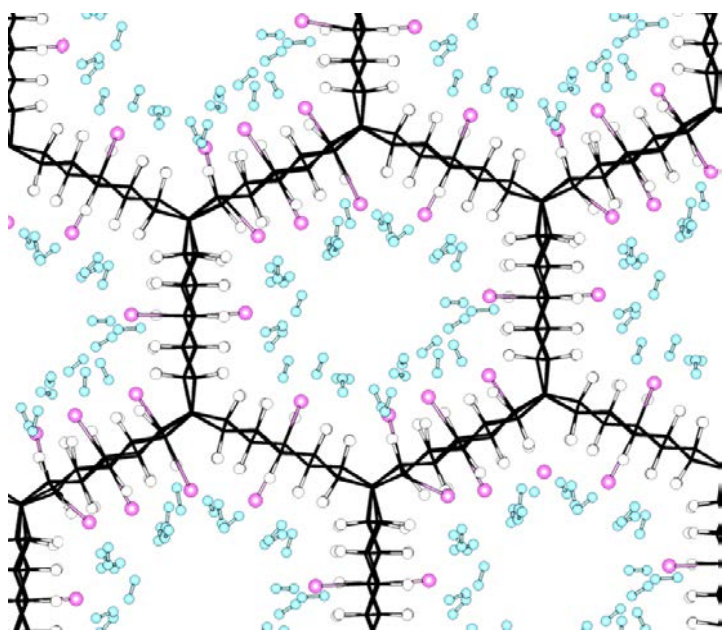

**Fig. 6.** An example of the porous framework structure based on the Li-Gr complex fragments organized in hexagonal cellular network with adsorbed hydrogen molecules (3 $H_2$ per 1 Li). The gravimetric and volumetric capacities of the structure are 6.07 wt% and 0.074 kg $H_2$/L, respectively. Carbon atoms are drawn by lines, lithium atoms are marked by red circles of larger radius and H-atoms are marked by cyan and white circles with smaller radius for the molecular and atomic hydrogen forms, respectively.

The important question is how the complexes can be structured in the bulk material. We proposed that the metal-graphane complex can be organized in the porous framework similar to graphene in cell configuration.[47-50] For example, the Li-Gr complex can be organized in the bulk structure presented in the Fig. 6. The gravimetric and volumetric capacities of such organometallic network can reach 6.1 wt% and 0.074 kg $H_2$/L, respectively which not only exceed the DOE 2009 requirement[51] (4.5 wt% $H_2$, 0.028 kg $H_2$/L) but also satisfy future target of DOE 2017 (5.5 wt% $H_2$, 0.040 g $H_2$/L).

## IV. CONCLUSIONS

We have studied organometallic complexes of graphane with adsorbed alkali metal atoms (Li, Na, K) as a promising material for hydrogen storage. The complexes with bonded lithium, sodium and potassium ions absorb 12.20 wt%, 10.33 wt% and 8.56 wt% of molecular hydrogen, respectively, with a binding energy of ~ -0.20 eV/$H_2$ molecule which greatly exceeds DOE requirement. The study of the thermodynamics of these systems demonstrates that the Li-Gr complex is the most promising candidate for hydrogen storage: at T = 300 K and P = 5 - 250 atm, the complex can absorb up to 3 $H_2$ molecules per metal atom with adsorption rate 9.44 wt%. Complexes with sodium and potassium atoms store hydrogen under more severe conditions. Finally, we have proposed the structure of porous framework based on the considered complexes with gravimetric and volumetric capacities satisfied DOE 2009 and 2017 targets.

## V. ACKNOWLEDGMENTS

We are grateful to the Joint Supercomputer Center of the Russian Academy of Sciences and 'Lomonosov' supercomputer of Moscow State University for the possibility of using a cluster computer for quantum-chemical calculations. This work was supported by JAEA Research fellowship (P.V.A.) and Russian Ministry of Education and Science (Contract No. 16.552.11.7014) (L.Y.A. and P.B.S.). P.V.A. and P.B.S. also acknowledge JSPS and JAEA ASRC and Molecular Spintronics Group for hospitality and fruitful collaboration.

* Corresponding author: pbsorokin@gmail.com